\documentclass[aps,pre,showpacs,twocolumn]{revtex4-1}

\usepackage{amsmath,amssymb}
\usepackage{graphicx}
\usepackage{color}

\newcommand{\xy}[1]{\textcolor{black}{ #1}}

\newcommand{\dg}{\dot\gamma}

\begin{document}

\title{Friction-induced shear thickening: a microscopic perspective} \author{Moumita
  Maiti} \author{Annette Zippelius}\author{Claus Heussinger} \affiliation{Institute for Theoretical
  Physics, Georg-August University of G\"ottingen, Friedrich-Hund Platz 1, 37077
  G\"ottingen, Germany}

\begin{abstract}
  We develop a microscopic picture of shear thickening in dense
  suspensions which emphasizes the role of frictional forces, coupling
  rotational and translational degrees of freedom. Simulations with
  contact forces and viscous drag only, reveal pronounced shear
  thickening with a simultaneous increase in contact number and energy
  dissipation by frictional forces. At high densities, when the
  translational motion is severely constrained, we observe liquid-like
  gear-states with pronounced relative rotations of the particles
  coexisting with solid-like regions which rotate as a whole. The
  latter are stabilised by frustrated loops which become more numerous
  and persistent with increasing pressure, giving rise to an
  increasing lengthscale of this mosaique-like structure and a
  corresponding increase in viscosity.
\end{abstract}


\date{\today}

\maketitle

\section{Introduction}



The phenomenon of shear thickening in dense suspensions of particles
has caught considerable attention
recently~\cite{JaegerReview,fall2008,pan2015,fall2015,brown2012,
  heussinger2013,fernandez2013,Mari2013,mari2014,Mari2015,cates2013}. It
is now becoming clear, through experiments and simulations, that next
to the well studied hydrodynamical thickening
mechanism~\cite{wagner2009,brady1985,cheng2011,bian2014} there is a
second mechanism based on the presence or absence of inter-particle
friction~\cite{brown2012,heussinger2013,fernandez2013,Mari2013}. Similar
effects are observed in dry powders of granular particles
\cite{Grob2014,otsuki2009,bi2011,sarkar2015}. Theoretically~\cite{cates2013},
it has been argued that thickening, i.e. the increase of viscosity
with stress or strainrate, is due to the fact that in frictional
systems the viscosity diverges at lower particle volume fractions than
in the corresponding frictionless systems. For given volume fraction
$\phi$, the viscosity of the frictional system is therefore higher
than in a frictionless reference system. Thus, by ``switching-on'' the
frictional interactions as the stress is increased, a thickening
regime occurs. Depending on how fast this switch happens, the
thickening may be continuous or discontinuous. In particle-based
simulations such a switch has been implemented in
Ref.~\cite{Mari2013,mari2014,Mari2015}. In these simulations
frictional interactions are only active, when the normal force, that
is pushing the particles together, exceeds a certain threshold
$f^\star$. The onset-stress for thickening then naturally follows as
$\sigma_{\rm on}=f^\star/d^2$, with $d$ the diameter of the particles.

From dimensional analysis it is clear that an additional force scale
$f^\star$ is necessary, if deviations from Newtonian (or Bagnold) flow
are to be seen~\cite{lois2007PREa}.  This force scale may be
implemented in various
ways~\cite{fernandez2013,Mari2013,cates2013,heussinger2013,Grob2014}. In
Refs.~\cite{heussinger2013,Grob2014} it is introduced by giving the
particles a finite stiffness. Subsequently, in these simulations,
friction is not ``switched-on'', but is always active, both in the
thickening regime above the onset stress and in the Newtonian regime
below this threshold. Thickening, then, is not a transition from
frictionless to frictional rheology. It rather points to different
ways frictional forces can act. Apparently, there is a qualitative
change in the way frictional inter-particle forces affect the rheology
at the onset of thickening. It is the goal of this article to provide
a microscopic (particle-based) perspective for this change.

\section{Simulation}

We study a two-dimensional binary mixture of $N$ soft spheres,
$N/2$ spheres of diameter $d$ and $N/2$ spheres of
diameter $1.4d$. The particle volume fraction is defined as
$\phi = \sum_{i=1}^{N}\pi R_{i}^2/L^2$, where $R_{i}$ is the
radius of a particle $i$, and $L$ is the length of the simulation
box. The system is sheared along the x-direction and Lees-Edwards
periodic boundary conditions are used.



Particles interact via a standard spring-dashpot
interaction~\cite{cundall1979}. Two particles $i,j$ interact when they
are in contact, i.e., when their mutual distance r is smaller than the
sum of their radii $R_i+R_j$. The normal component of the interaction
force is $F_n = k_n [r-(R_i+R_j)]-\gamma_n \delta v_n$, where $k_n$ is
the spring constant, $\gamma_n$ the dashpot strength, and $\delta v_n$
the relative normal velocity of the two contacting particles.
\xy{While we include this dissipative coupling $\gamma_n$ in the
  simulations, it turns out to not play any role for the rheological
  behavior (see below).}  The tangential force component is
$F_t = k_t \delta_t$, with $\delta_t$ the tangential (shear)
displacement since the formation of the contact. The tangential spring
mimics sticking of the two particles due to dry friction. These
frictional forces are limited by the Coulomb condition
$F_t\leq \mu F_n$, with a constant, i.e., velocity independent
friction coefficient $\mu$.  \xy{We do not include a dissipative
  coefficient in the tangential direction. Still, there is tangential
  dissipation when contacts are sliding, via the Coulomb
  criterium. This choice of parameters is used primarily for
  simplicity in order to reduce the number of parameters. Still the
  essential physics is retained, namely the presence of tangential
  forces and dissipation.}

\xy{The system is subject to shear in the $x$-direction. We impose a
  fixed shear rate $\dot\gamma$, which is the key control parameter of
  the simulation}. Newton’s equations of motion
$m {\dot{\vec v}}_i = {\vec F}^{\rm cont}_i + {\vec F}^{\rm visc}_i$
are integrated with contact forces as specified above and a viscous
drag force, which implements the shear flow. The drag force
${\vec F}^{\rm visc}(\vec{v}_i) = -\zeta {\vec{\delta v}}_i$ is
proportional to the velocity difference
$\vec{\delta v}_i = {\vec v}_i-{\vec v}_{flow}$ between the particle
velocity ${\vec v}_i$ and the flow velocity
${\vec v}_{flow}(r_i) = {\hat e}_x\dot\gamma
y_i$~\cite{olssonPRL2007,durianPRL1995}.  The dissipative coefficient
$\zeta$ represents the viscosity of the surrounding fluid,
$\zeta\sim\eta_f$.  Inertial forces, while included in the equation of
motion, are not important for the small strainrates considered, and
only affect the flow at larger strainrates. \xy{The inertial number is
  generally smaller than $O(10^{-3})$.}

A key difference between frictionless and frictional particles is the
presence of rotational degrees of freedom. Indeed, we will argue below
that the coupling of rotation and translation is an important factor
in understanding the rheological properties. The equation of motion
for the rotational velocity ${\vec \omega}_i$ of particle $i$ is
$I{\dot{\vec \omega}}_i = {\vec T}_i^{\rm cont}$, where the torques
derive from the contact forces given above. The torque from the
\xy{viscous drag force is neglected for simplicity}.

As units we use the spring constant $k$, the particle diameter $d$ as
well as the mass density $\rho$. System size is $N = 10000$ for most
densities except at lower densities it is $2500$. Friction coefficient
and dissipative coefficients are $\mu = 1$ and $\zeta = \gamma_n = 0.1$.


\section{Results}

\begin{figure}[ht]
  \includegraphics[angle=0, width=0.9\linewidth,clip=true]{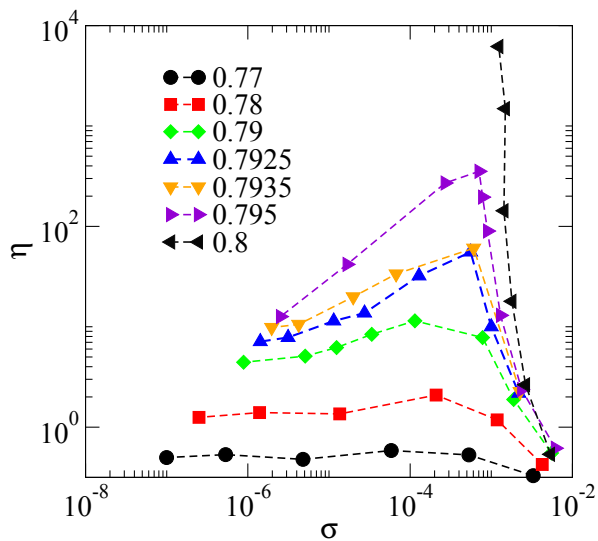}
  \caption{\label{fig:flowcurve} Viscosity $\eta$ vs. stress $\sigma$
    for different volume fractions $\phi$.}
\end{figure}

The stress tensor is calculated from the virial expression, for
example for the shear stress,
\begin{eqnarray}\label{eq:}
\sigma = \frac{1}{2L^2}\sum_i F_i^xy_i
\end{eqnarray}
where the sum is taken over all particles $i$. From this the viscosity
is defined as $\eta=\sigma/\dot\gamma$. This is plotted in
Fig.~\ref{fig:flowcurve} as a function of stress $\sigma$ for various
volume fractions $\phi$.

For the smallest $\phi$ probed the viscosity is constant at small and
intermediate stresses and then decreases. The system is shear
thinning. This is also the behavior observed in a ``reference''
system, where frictional forces have been switched
off~\cite{olssonPRL2007}. At slightly higher volume fractions, there
is a stress value, $\sigma_{\rm on}\approx 10^{-6}k$, above which the
system displays shear thickening. The precise value for this onset
stress is difficult to obtain, as the associated strainrates are very
small and simulations get prohibitively long. Such a thickening regime
is absent in the frictionless reference system. At still higher $\phi$
the viscosity is infinite below the yield stress
$\sigma_y\approx 10^{-3}k$ and the system is solid there. The small
prefactor for the onset of thickening already indicates that this
phenomenon is not straightforwardly related to the direct particle
interactions but is rather a many-body network effect.

From the diagonal elements of the stress tensor, i.e. the pressure
$p$, we obtain a similar behavior. The (macroscopic) friction
coefficient $\mu_m=\sigma/p$ only slightly varies between $0.29$ and $0.31$.




\begin{figure}[ht]
  \includegraphics[angle=0, width=0.9\linewidth,clip=true]{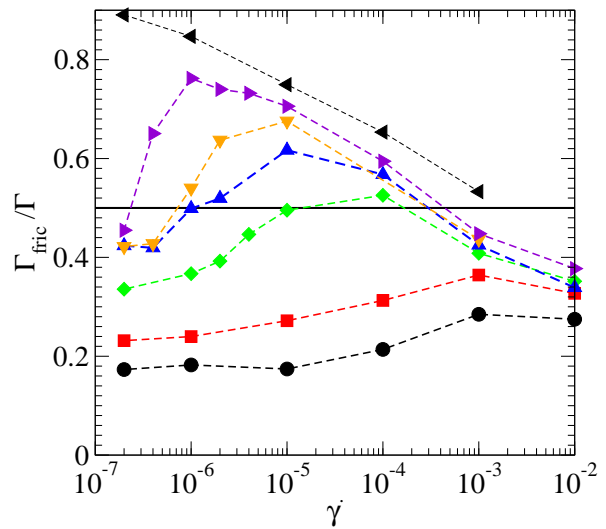}
  \caption{\label{fig:ratio} Fraction of energy dissipated via
    frictional interactions $\Gamma_{\rm fric}/\Gamma$ vs. strainrate $\dg$ for different volume
    fractions $\phi$. Color code as in Fig.~\ref{fig:flowcurve}.}
\end{figure}

While the crucial role of friction for the shear thickening behavior
is demonstrated from the comparison with the frictionless reference
system, we can quantify this influence by calculating the amount of
energy dissipation that is due to frictional forces. The \xy{viscosity
  $\eta$ is quite generally connected} to the total energy dissipation
rate $\Gamma=L^2\eta\dot\gamma^2$. To determine the contribution
$\Gamma_{\rm fric}$ from frictional forces we calculate the viscous
contribution as~\cite{PhysRevLett.109.105901}
\begin{eqnarray}
  \label{eq:Gamma_v}
  \Gamma_{\rm visc} = \zeta N\langle\delta v^2\rangle
\end{eqnarray}
where $\delta v$ is the non-affine particle velocity $v-v_{\rm flow}$,
i.e. after the flow velocity is subtracted. The frictional
contribution then is $\Gamma_{\rm fric}= \Gamma - \Gamma_{\rm visc}$
(the dissipation from inelastic collisions is very small and can be
neglected). The ratio $\Gamma_{\rm fric}/\Gamma$ is plotted in
Fig.~\ref{fig:ratio} for the same data as in
Fig.~\ref{fig:flowcurve}. Most notably is the strong increase of the
frictional contribution in the thickening regime. However, friction
also plays a non-negligible role when no thickening is observed (at
lower $\phi$ and small strainrates).

\begin{figure}[ht] 
   \centering
   \includegraphics[width=0.9\linewidth]{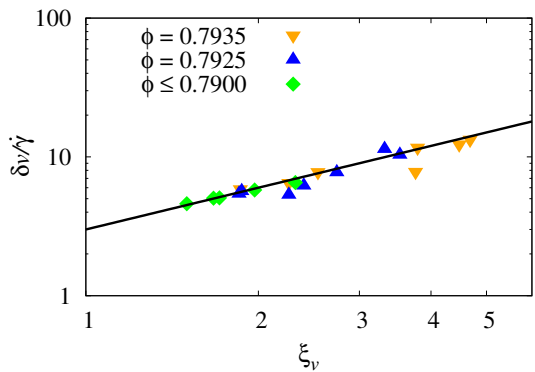}
   \caption{\xy{Normalized translational velocity $\delta v/\dg$
       vs. correlation lengthscale $\xi_v$. Velocity is proportional
       to length scale, $\delta v\approx 1.7\xi_v\dot\gamma$.}}
   \label{fig:velocity_xi}
 \end{figure}
\begin{figure}[ht] 
   \centering
   \includegraphics[width=0.9\linewidth]{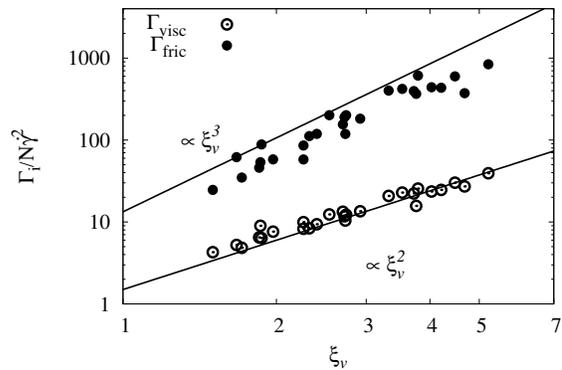}
   \caption{\xy{Normalized energy dissipation $\Gamma_i/N\dg^2$
       vs. lengthscale $\xi_v$, where $i$ stands for viscous (visc) and
       frictional (fric) contributions, respectively. The frictional contribution is
       multiplied by ten to avoid overlapping data points. Data taken
       from different strainrates, volume fractions and system
       sizes.}}
   \label{fig:dissipation_xi}
 \end{figure}

In previous work~\cite{heussinger2013} we have argued that the viscous
dissipation induced by particle translational velocities is coupled to
a length-scale that occurs in the spatial correlation of velocities
$C_v=\langle \delta v(x)\delta v(0)\rangle$. The long-range part of
the correlation function decays exponentially, $C_v\sim e^{-x/\xi_v}$,
giving rise to a length-scale $\xi_v(\dot\gamma)$ that was shown to
increase in the thickening regime~\xy{\cite{heussinger2013,footnote}}.
\xy{Examples of this correlation function are displayed in Fig.~4 of
  Ref.~\cite{heussinger2013}.
  Here we furthermore show that this correlation length $\xi_v$ is
  tightly connected to the fluctuations of the particle velocity
  $\delta v\equiv \sqrt{\langle \delta v_i^2\rangle}$.  Fig.~\ref{fig:velocity_xi} emphasizes that both
  quantities are proportional to each other, with a proportionality
  constant of order unity, $\delta v\approx 1.7\xi_v\dot\gamma$.
  This indicates that typical velocities are no longer set by the size
  of the particles as $v\sim d\dg$, rather the fundamental
  length-scale is now the correlation length $\xi_v$, which takes the
  meaning of an effective particle size.
  It needs to be emphasized that this proportionality is a nontrivial
  result, as it represents a direct link between a single-particle
  observable (the velocity) and a two-point correlation function.
  As, according to Eq.~(\ref{eq:Gamma_v}), the velocities enter the
  viscous dissipation rate $\Gamma_{\rm visc}\sim \delta v^2$, we need to have
  $\Gamma_{\rm visc}\sim \xi_v^2\dot\gamma^2$, for consistency.
This is indeed observed in Fig.~\ref{fig:dissipation_xi} (open symbols).
Interestingly, the frictional contribution to the energy dissipation
rate $\Gamma_{\rm fric}$ scales \xy{differently,}
$\Gamma_{\rm fric}\sim \xi_v^3\dg^2$ (filled symbols in
Fig.~\ref{fig:dissipation_xi}). The additional factor of $\xi_v$ as
compared to $\Gamma_{\rm visc}$ can be understood by noting that the
frictional dissipation can be written as
$\Gamma_{\rm fr}\sim (\mu f_n)\cdot \delta v$, which is just the
dissipation of a sliding contact, sliding at velocity
$\delta v\sim \xi_v\dg$ under normal force $f_n$. Assuming that
predominantly viscous forces lead to the build-up of the normal force,
we can write $f_n\sim \Gamma_{\rm visc}/\dg\sim \xi_v^2\dg$ and,
consequently, $\Gamma_{\rm fric}\sim \xi_v^3\dg^2$.  }

\begin{figure}[ht]
  \includegraphics[angle=0, width=0.7\linewidth,clip=true]{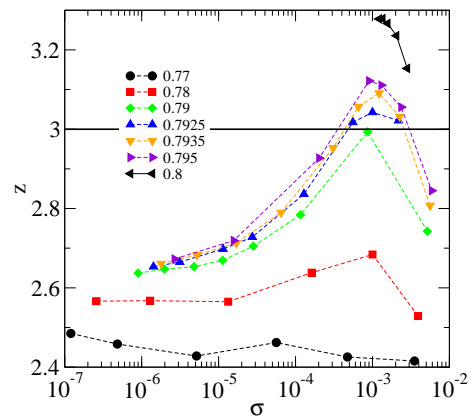}
  \caption{\label{fig:contacts} Average number of contacts per
    particle $z$ vs. stress for different volume fractions. At the
    onset of thickening the connectivity starts to rise and peaks at
    the yield-stress with $z\approx z_{\rm iso}$.}
\end{figure}

In order to better understand the microscopic origins of these
phenomena we first take a look at the average number of contacts per
particle, $z$ (see Fig.~\ref{fig:contacts}). In the zero-stress limit
the connectivity is below the isostatic value of $z_{\rm iso}=3$, as
expected for a fluid. The isostatic point represents the minimal
connectivity value at which a jammed state can first be
formed. Typically, however, and depending on the preparation
procedure, frictionally jammed states have a somewhat higher
coordination, $z_{\rm jam}>z_{\rm
  iso}$~\cite{agnolin2007PRE,song2008Nature}. 
%
With increasing stress, in the thickening regime, also $z$
increases. At the yield-stress $\sigma_y$ our system is approximately
isostatic, before the connectivity decreases again in the regime of
shear-thinning. The latter effect is also observed in the frictionless
reference system, however, the increase of $z$ with stress (or
pressure) is special for the frictional system.

In the thickening regime, we also observe a universal functional
dependence $z(\sigma)$, as the connectivity only weakly depends on
volume fraction $\phi$. One may therefore identify the connectivity as
the relevant field that triggers the onset of thickening.

\begin{figure}[h]
  \includegraphics[clip=true,angle=0, width=0.7\linewidth]{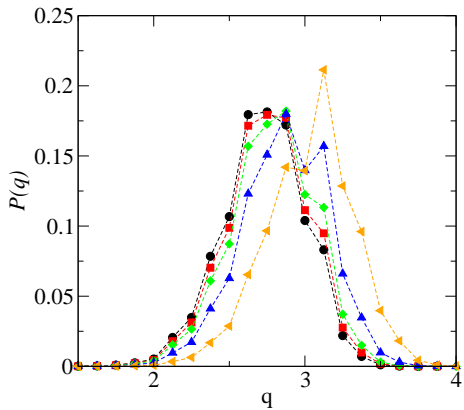}
  \caption{\label{fig:distribution} Probability distribution $P(q)$ of
    the coarse-grained contact number $q$, for the volume fraction
    $\phi=0.7935$ and different strainrates $\dot\gamma=2\times
    10^{-7}$ (black circles), $4\times 10^{-7}$ (red squares),
    $10^{-6}$ (green diamonds), $2\times 10^{-6}$ (blue triangles up),
    $4\times 10^{-6}$ (orange triangles left).}
\end{figure}

More detailed information about the inter-particle contacts is given
by the probability distribution for $z$. To this end we use a slightly
coarse-grained version $q$ of the connectivity. To calculate $q$, the
particle based contact number $z$ is averaged over a circular region
of radius $R=3$. This value for the coarse-graining parameter
corresponds roughly to the second minimum in the pair correlation
function. In Fig.~\ref{fig:distribution} the probability distribution
$P(q)$ is shown for different strainrates and $\phi = 0.7935$.  Most
notably, a distinctive shoulder at $q\approx z_{\rm iso}=3$ develops
in the thickening regime. Thus, thickening is associated with the
formation of local patches of elevated connectivity.

This is in close analogy to the recent work of Henkes {\it et al.}
\cite{PhysRevLett.116.028301}, that observe a distribution of rigid
clusters, the typical size of which diverges when crossing the jamming
transition. It is tempting to identify these clusters with our regions
of elevated connectivity, as well as their size with the above
introduced length-scale $\xi_v$.

Shear thickening then corresponds to the formation of temporarily
rigid clusters in a see of floppy particles. The role of pressure is
to stabilize these clusters as frictional contacts get more stable for
higher pressures. As pressure increases with increasing strainrate,
the size of these regions grows larger, and dissipation is increased
via the contributions presented in Fig.~\ref{fig:dissipation_xi}. All
this is not observed in frictionless systems. Indeed,
Ref.~\cite{PhysRevLett.116.028301} claims that no diverging cluster
size is observed without friction.

\begin{figure}[ht] 
   \centering
   \includegraphics[width=0.9\linewidth]{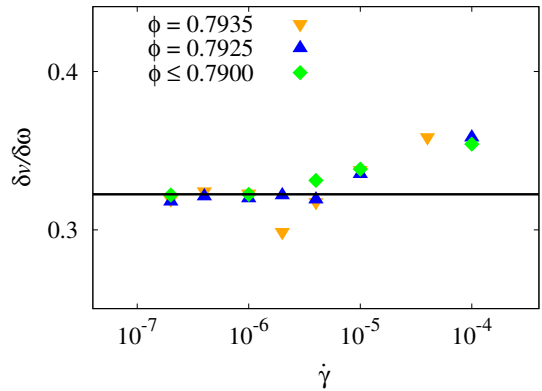}
   \caption{\xy{Ratio of translational to rotational velocity
       vs. strainrate $\dg$. The ratio is nearly constant,
       $\delta v/\delta\omega\approx 0.32d$; note the linear y-axis.}}
   \label{fig:vel_rot_coupling}
\end{figure}

\begin{figure}[ht] 
   \centering
   \includegraphics[width=0.9\linewidth]{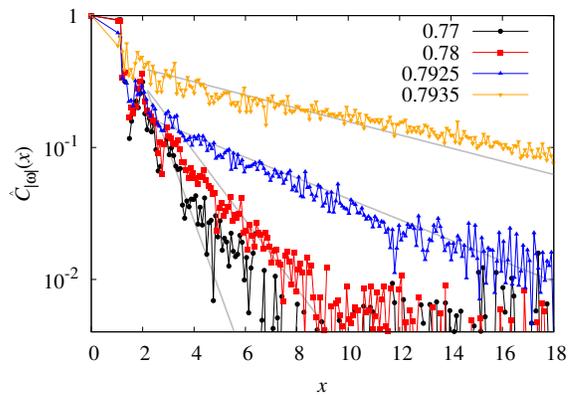}
   \caption{\xy{Normalized correlation function
     $\hat C_{|\omega|}(x)=C_{|\omega|}(x)/\langle
     \delta|\omega|^2\rangle$ vs. $x$ for $\dg=10^{-6}$ and different
     $\phi=0.77,0.78,0.7925,0.7935$. The grey lines are fits using an
     exponential function $\sim \exp(-x/\xi_\omega)$, where
     $\xi_\omega=2.3\xi_v$.}}
   \label{fig:absomega_corr}
\end{figure}

A key difference between frictional and frictionless particles is the
presence of rotational degrees of
freedom. Fig.~\ref{fig:vel_rot_coupling} indicates that particle
rotations are strongly coupled to translations. The root-mean square
fluctuations of the rotational velocity $\delta\omega$ is always about
three times the translational velocity $\delta v$.
\xy{With this one may anticipate that rotational correlations are also
  coupled to translational correlations, i.e. characterized by a
  similar lengthscale as $\xi_v$ introduced above. To test this
  assumption we define
  $C_{|\omega|}(x) = \langle
  \delta|\omega(x)|\delta|\omega(0)|\rangle$ for the correlation of
  the absolute values of particle rotations.
  Fig.~\ref{fig:absomega_corr} indicates that this function acquires a
  long-range decay in the shear-thickening regime. While the decay is
  not strictly exponential, we have tried to fit a function of the
  form $\sim\exp(-x/\xi_\omega))$, where $\xi_\omega =c\xi$ with a
  universal constant $c\approx 2.3$. This procedure works very well so
  that we can conclude that the correlations of translational and
  rotational velocities are indeed governed by the same lengthscale
  $\xi\propto \xi_v\propto \xi_\omega$.  }

To understand the reason and the consequences of the coupling between
rotations and translations consider the situation that two
individually rotating particles come into contact. Once the contact is
formed, the frictional forces will act to reduce the relative
tangential motion of the two particles at the point of contact, up
until the condition $v_1-v_2=(d_1\omega_1+d_2\omega_2)/2$ is met, when
there is no relative motion at the point of contact.

One may think of two extreme scenarios to achieve this goal: either
rotations adapt (via frictional dissipation) without generating
relative translational motion, $v_1\approx v_2$. In this case
particles will rotate at the same speed in opposite sense, i.e.
$\omega_1=-\omega_2$ (equal-sized particles). Alternatively,
rotational velocity may be transformed in translational velocity,
$v\sim d\omega$. In this case particles will rotate around each other
instead of around their respective particle centers. Such a motion
leads to viscous dissipation.

\begin{figure}[h] 
   \centering
   \includegraphics[width=2.5in]{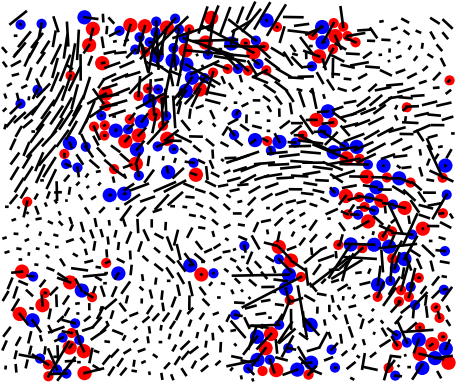}
   \caption{Snapshot of a small part of the system at
     $\dot\gamma=10^{-6}$ and $\phi=0.7925$. Color code represents the
     rotational state of the particles (only strongly rotating
     particles are shown): large clockwise rotation (red), large
     anti-clockwise (blue). Lines are the non-affine velocities of the
     particles. Clusters and vortices seem to form in between clusters
     of strongly rotating particles.}
   \label{fig:cluster_gear}
\end{figure}

In general contacts do not come in pairs, and particles are tightly
constrained by the surrounding network of other contacts. We have
quantified this influence via the lengthscale $\xi$ that governs the
correlations of particle velocity. In extension to the two-particle
picture of contact formation we can therefore think of a patch of
particles of linear size $\xi$. In this patch, particles are either
(in the extreme cases just discussed) rotating coherently without
relative slip and with only little movement in space, or they are
translating more or less as solid blocks forming clusters or vortices
\cite{olssonPRL2007,heussingerEPL2010,tanguyPRB2002}. Such a behavior
can be seen in Fig.~\ref{fig:cluster_gear}. The state of rotation of
particles is displayed together with their nonaffine velocities. Most
prominent are the vortex- or block-like structures in places where the
rotational velocities are small (particles are only shown when their
rotational velocity is large.)

\begin{figure}[h] 
   \centering
   \includegraphics[width=2.5in]{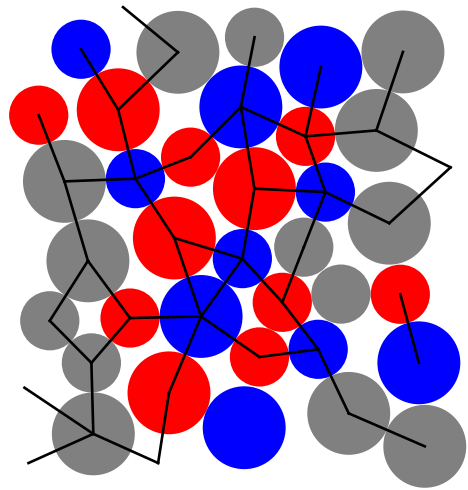}
   \caption{Snapshot of a small part of the system at
     $\dot\gamma=10^{-6}$ and $\phi=0.7925$. Color code represents the
     rotational state of the particles: small rotation (gray), large
     clockwise (red), large anti-clockwise (blue). Contacts are
     indicated by lines between particles. Note the alternating color
     of connected particles highlighting the strong rolling motion.}
   \label{fig:bearing_structure}
\end{figure}

\begin{figure}[h] 
   \centering
   \includegraphics[width=2in,clip=true]{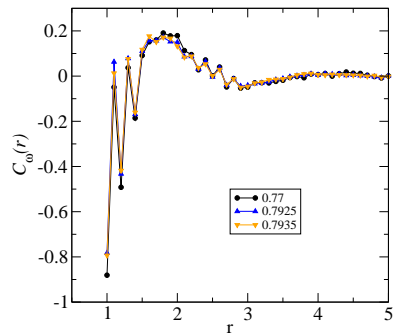}
   \caption{Normalized correlation function
     $\hat C_\omega(r) = \langle
     \delta\omega(0)\delta\omega(r)\rangle/\langle
     \delta\omega^2\rangle$ (no absolute values as compared to
     Fig.~\ref{fig:absomega_corr}) vs. distance $r$. Strong
     anti-correlation of contacting particles (those at
     $r=1.0, 1.2, 1.4$).}
   \label{fig:rotational_correlation}
\end{figure}

The state of coherent rotations without slip has been called a
bearing-state~\cite{astrom2000,alonso2006}. Because of the
counter-rotation of neighboring particles the typical particle
rotation $\delta\omega\sim \dot\gamma(\xi/d)$ in such a state is much
larger than the average rotation, which can be taken to be imposed by
the strainrate of deformation, $\bar\omega\sim
\dot\gamma$~\cite{halsey2009}.
Such a counter-rotation is readily observed in the exemplary shnapshot
of Fig.~\ref{fig:bearing_structure}, where connected particles
typically show alternating sign of their rotational velocity. A
similar conclusion is obtained from the correlation function of
particle rotations,
$C_\omega(r) = \langle \delta\omega(0)\delta\omega(r)\rangle$ (without
absolute values) as displayed in
Fig.~\ref{fig:rotational_correlation}. Contacting particles (those at
distances, $r=1.0, 1.2, 1.4$) have a strongly negative correlation of
their rotational velocities.

However, not all contact networks allow for bearing states. A crucial
role is played by contact loops. A contact loop is a set of contacting
particles that form an ``empty'' loop, i.e. with no other (contacting)
particle inside the loop. A $k$-loop then is a contact loop with $k$
particles. Loops with an odd number of particles, in particular
three-particle loops, play an important role in the process of
rotational frustration~\cite{tordesillas2012}. These loops do not
allow for coherently rotating states and rotation is fully frustrated,
similar to the spins in antiferromagnetic triangular plaquettes.

It has been argued that the presence of these loops stabilize the
contact network in packings of granular particles, for example, by
inhibiting the buckling of force
chains~\cite{tordesillas2010,astrom2000}. In a dynamic setting during
shear flow there is a constant breaking-up of smaller loops into
larger ones, and the reverse process of collapse of larger loops into
smaller ones. The presence of 3-loops then inhibits gear states and
favors the formation of blocks or vortices. This increases energy
dissipation and leads to the shear-thickening state.

\begin{figure}[ht]
  \includegraphics[angle=0,clip=true,width=0.9\linewidth]{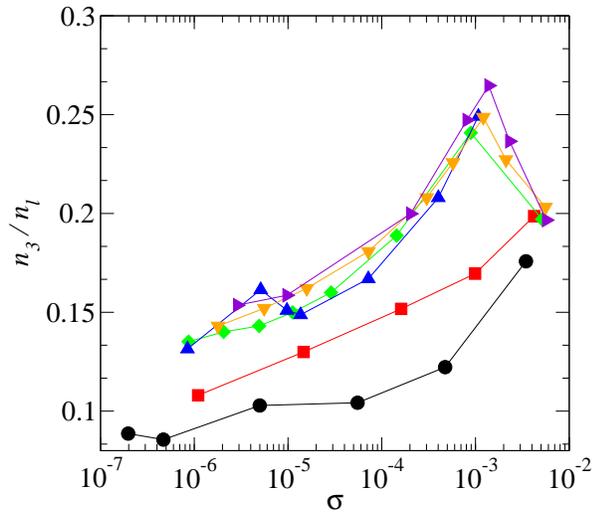}
  \caption{\label{fig:3loops} Fraction $n_3/n_l$ of 3-particle loops
    vs. stress $\sigma$ for different $\phi$. Color coding is as in
    Fig.~\ref{fig:flowcurve}.}
\end{figure}

From the Euler characteristic the total number of loops per particle
only weakly increases with the connectivity as $n_l = (z/2-1)$. By way
of contrast, as is evidenced by Fig.~\ref{fig:3loops} the number of
3-loops strongly increase when entering the thickening regime --
roughly by a factor of three.
%

One can therefore think of a 3-loop as a defect with the lengthscale
$\xi$ playing the role of the defect size. Once contact rotation
becomes frustrated the rotational velocity of the individual particles
can be turned into translational velocity of a structure of size
$\xi$, the velocity scaling as $\delta v\sim \dot\gamma\xi$. The viscous
dissipation then follows as $\Gamma_v\sim \zeta \delta v^2\sim \xi^2$.

\section{Conclusion}
\label{sec:discussion}

We study the phenomenon of shear thickening in dense athermal
suspensions. The mechanism responsible for shear thickening is agreed
to consist of a transition from frictionless to frictional
rheology. However, it is not clear how this transition comes about on
the microscopic level of particles. In this work we have started to
fill this gap of understanding. The goal is to provide particle-based
explanations of how structural, dynamical and rheological properties
change in the shear-thickening regime at high particle volume
fractions.  In line with recent work we have observed that shear
thickening is associated with increased frictional interactions. We
have quantified this importance by measuring the amount of energy that
is dissipated via frictional forces. Nevertheless, frictional
dissipation is also present in the Newtonian regime and can never be
neglected.  This points to different ways frictional forces may act.
A clue to this variability is provided by the presence of the
correlation lengthscale $\xi$, that shows up in the spatial
correlations of particle translational and rotational velocities, and
which starts to grow at the onset of thickening. Subsequently, the
lengthscale governs viscous and frictional energy dissipation as
$\Gamma_{\rm visc}\sim \xi^2\dg^2$ and
$\Gamma_{\rm fric}\sim \xi^3\dg^2$, respectively.  On the kinematic
side, the lengh-scale sets the scale for the particle velocities, the
fluctuating (non-affine) contributions scaling as
$\delta v\sim \xi\dg$. The key difference between frictionless and
frictional systems is the presence of rota- tional degrees of
freedom. We have found that rotational velocities are strongly coupled
to translations, their ratio being nearly constant
($\delta v/\delta\omega\approx 0.32$) throughout the thickening
regime. Thus, also rotational velocities scale with the correlation
lengthscale, $\delta \omega \sim (\xi/d)\dg$.

We have interpreted these findings with the help of the concept of
rotational frustration, that occurs whenever new contacts lead to
loops with an odd number of parti- cles. These loops block the
coherent rotation of a particle patch and favor the interchange of
rotational into trans- lational velocities. In consequence, the
particles start to move as (temporarily) rigid clusters and viscous
energy dissipation is enhanced. 

Pressure is expected to play an important role in these processes. At
small pressures these structures quickly relax, and the resulting
shear stability is weak. At higher pressures, in the thickening
regime, these structures persist resulting in a higher stability.  In
this picture the system consists of a mosaique of solid-like blocks or
vortices (maybe formed by frustrated loops) and liquid-like
gear-states (see Fig.~\ref{fig:bearing_structure}). Vortex particles
do not rotate, but vortices rotate as solid bodies. Gear particles, in
contrast, strongly rotate relative to each other. The size of these
structures increases as pressure increases, and so does the viscosity.
There are striking similarities with the recent work of Henkes {\it et
  al.}\cite{PhysRevLett.116.028301}. By implementing a rigid cluster
decomposition they detect temporarily rigid particle patches, the
typical size of which diverges when crossing the jamming
transition. In that work the approach to jamming is associated with
increasing volume fraction, while here we deal with (presumably) the
same phenomenon as a function of strainrate to explain the
shear-thickening state. More work will be necessary to fully work out
these analogies.

\begin{acknowledgments}
  We acknowledge financial support by the German Science Foundation
  via the Emmy Noether program (He 6322/1-1).
\end{acknowledgments}


\end{document}